\documentclass[conference]{IEEEtran}
\IEEEoverridecommandlockouts
\usepackage{cite}
\usepackage{amsmath,amssymb,amsfonts}
\usepackage{algorithmic}
\usepackage{graphicx}
\usepackage{textcomp}
\usepackage{xcolor}
\def\BibTeX{{\rm B\kern-.05em{\sc i\kern-.025em b}\kern-.08em
    T\kern-.1667em\lower.7ex\hbox{E}\kern-.125emX}}

\begin{document}

\title{Photon Efficiency Limits in the Presence\\ of Background Noise
\thanks{This work is part of the project ``Quantum Optical Communication Systems'' carried out within the  TEAM programme of the Foundation for Polish Science co-financed by the European Union under the European Regional Development Fund.}
}

\author{\IEEEauthorblockN{K. Banaszek, W. Zwoli\'{n}ski, L. Kunz, M. Jarzyna}
\IEEEauthorblockA{\textit{Centre for Quantum Optical Technologies, University of Warsaw} \\
Banacha 2c, 02-097 Warsaw, Poland \\
\{k.banaszek, w.zwolinski, l.kunz, m.jarzyna\}@cent.uw.edu.pl}
}

\maketitle

\begin{abstract}
We identify theoretical limits on the photon information efficiency (PIE) of a deep-space optical communication link constrained by the average signal power and operated in the presence of background noise. The ability to implement a scalable modulation format, Geiger-type direct photon counting detection, and complete decoding of detection events are assumed in the analysis. The maximum attainable PIE is effectively determined by the background noise strength and it exhibits a weak, logarithmic dependence on the detected number of background photons per temporal slot.
\end{abstract}


\section{Introduction}
In the photon-starved regime, typical to operation of deep-space optical communication links \cite{BiswasICSOS2017,SodnikICSOS2017},
a relevant figure of merit is the {\em photon information efficiency} (PIE) quantifying the amount of information that can be transmitted per photon at the carrier frequency. The purpose of this contribution is to identify theoretical limits on the attainable PIE when the received signal is accompanied by background noise. We find that when comprehensive capabilities to optimize the modulation format and to decode detection outcomes are assumed, one can in principle  maintain a non-zero PIE value  with the vanishing average signal power at a given background noise level. This remains the case also when the signal-to-noise ratio (SNR) becomes negative when expressed in dB units. Importantly, maintaining non-zero PIE makes the information rate scale as $r^{-2}$ with the link range $r$. This is in contrast with the commonly held view that the information rate of deep-space optical communication links exhibits unfavorable $r^{-4}$ scaling once the received signal becomes dominated by the background noise \cite{Moision2014}, which would imply the need to revert to radio frequency (RF) communication for very deep space missions.

\section{Information Rate}

It is instructive to start by reviewing the relation between the information rate $R$ and the PIE. It will be convenient to introduce the average number of signal photons $n_a$ and background photons $n_b$ detected over the duration of an elementary temporal slot, given by the inverse of the modulation bandwidth $B$. These two quantities will be used to parameterize the photon information efficiency $\textrm{PIE}(n_a,n_b)$. The explicit expression for $n_a$ reads
\begin{equation}
n_a =  \eta_{\textrm{rx}}\cdot \eta_{\textrm{ch}} \cdot \frac{P_{\textrm{tx}}}{Bhf_c}.
\label{Eq:na}
\end{equation}
The meaning of mathematical symbols used in this section is specified in Table~\ref{Tab:Symbols}. The link range $r$ enters explicitly the expression for the channel transmission $\eta_{\textrm{ch}}$ given by
\begin{equation}
\eta_{\textrm{ch}} = \frac{1}{r^2} \cdot  f_c^2 \cdot \frac{\pi^2 D_{\textrm{rx}}^2 D_{\textrm{tx}}^2}{16c^2},
\label{Eq:etach}
\end{equation}
assuming that attenuation of the propagating signal is due to diffraction only.

\begin{table}[t]
\caption{Meaning of mathematical symbols}
\begin{center}
\begin{tabular}{|c|l|}
\hline
Symbol & \multicolumn{1}{c|}{Meaning}\\
\hline\hline
$\textrm{PIE}$ & photon information efficiency \\
\hline
$R$ & information rate \\
\hline
$n_a$ & average detected signal photon number per slot \\
\hline
$n_b$ & average detected background photon number per slot \\
\hline
$\eta_{\textrm{rx}}$ & receiver efficiency \\
\hline
$\eta_{\textrm{ch}}$ & channel transmission \\
\hline
$P_{\textrm{tx}}$ &  signal power radiated by the transmitter\\
\hline
$B$ & modulation bandwidth \\
\hline
$h$ & Planck's constant \\
\hline
$f_c$ & carrier frequency \\
\hline
$r$ & link range \\
\hline
$D_{\textrm{rx}}$ & receiver antenna diameter \\
\hline
$D_{\textrm{tx}}$ & transmitter antenna diameter \\
\hline
$c$ & speed of light in vacuum \\
\hline
\end{tabular}
\label{Tab:Symbols}
\end{center}
\end{table}

Using (\ref{Eq:na}) and (\ref{Eq:etach}), the information rate $R = B \cdot n_a \cdot \textrm{PIE}$ can be written as
\begin{equation}
R =
  \frac{1}{r^2} \cdot f_c \cdot P_{\textrm{tx}} \cdot
 \textrm{PIE}(n_a,n_b)  \cdot
 \frac{\pi^2 \eta_{\textrm{rx}} D_{\textrm{rx}}^2 D_{\textrm{tx}}^2  }{16 h c^2}.
\end{equation}
The above expression applies to both RF and optical links.
The $r^{-2}$ factor stems from the attenuation of the received signal power with the distance.
The factor linear in the carrier frequency $f_c$ is the result of an interplay between diffraction losses that become reduced at optical frequencies due to the shorter carrier wavelength and the energy of a single photon that is higher in the optical band. The overall scaling of the information rate with the distance is determined by the behavior of $\text{PIE}(n_a,n_b)$ with the increasing range $r$ which enters through the first argument $n_a$ as implied by the definitions (\ref{Eq:na}) and (\ref{Eq:etach}).

For conventional coherent detection, PIE is limited by $(2\log_2\mathrm{e})/(1+n_b)$ bits/photon, with the maximum value reached when $n_a \rightarrow 0$. In RF communication the detected background is at least at the level of tens of photons, rendering PIE a fraction of one. Conversely, in the optical band one usually has $n_b \ll 1$, which enables nearly shot-noise limited coherent detection \cite{GuntKhanOPT2017}. Furthermore, the ability to count single photons at optical frequencies facilitates implementation of high-PIE modulation formats that surpass the coherent detection limit in the photon-starved regime. However, it has been recognized for a long time \cite{GordonProcIRE1962} that the information efficiency of photon counting is limited by the noise present in the channel.

\section{Model}

The model considered here is based on direct detection of a scalable $M$-ary modulation format that uses $M$ symbols over a bandwidth $B$. One symbol occupies a time frame with an overall duration $t_s = M \cdot B^{-1}$. The generic example of such a format is pulse position modulation (PPM), where the entire optical energy is concentrated in one of $M$ elementary temporal slots of duration $B^{-1}$. There exists a variety of modulation formats with equivalent efficiency, such as frequency shift keying (FSK), or Hadamard words composed from the binary phase shift keying (BPSK) constellation \cite{GuhaPRL2011,BanaszekICSOS2017}. From the physical perspective, a unifying feature of all these formats is that the entire optical energy $n_s = Mn_a$ of a symbol is carried by one of $M$ orthogonal modes extending over duration $t_s$ \cite{BanaszekJachuraICSO2018}. The modes are separated at the receiver. The separation is realized in the temporal domain for the PPM format, in the spectral domain for the FSK format, or using an interferometric setup for BPSK Hadamard words. The symbol is read out by counting photons separately in each orthogonal mode and identifying the signal mode carrying the optical energy. Without background noise, the principal impairment is that none of the modes within a given frame generate a photocount, resulting in the erasure of the input symbol. In the noiseless case, the PIE of an $M$-ary modulation format approaches $\log_2 M$ when $n_s \ll 1$ and it can be made arbitrarily high if there are no restrictions on the format order $M$.

In the presence of background noise, photocounts may be registered for multiple modes occupying one time frame. The most powerful complete decoding strategy is to recover the information from all combinations of photocount events that could have occurred within a frame. For such a scenario, the PIE is lower bounded by a compact expression based on relative entropy \cite{HamkinsISIT2004}. In the following, we will assume Geiger-type photon counting which discriminates only whether at least one photon or none at all have been registered. The photocount probability for the mode carrying the symbol optical energy reads $p_c= 1 - \exp(- n_s - n_b)$ while for other modes it is equal to $p_b = 1 - \exp(-n_b)$, and the bound takes the form
$
\textrm{PIE} \ge n_s^{-1} \cdot D \bigl( p_c
\bigr|\bigl| M^{-1} p_c + (1-M^{-1}) p_b \bigr)
$
where $D(\, \cdot \, || \, \cdot \, )$ 
denotes the relative entropy (Kullback-Leibler divergence) between two binary probability distributions.

\begin{figure}[!ht]
\begin{centering}
\includegraphics[width=0.83\columnwidth]{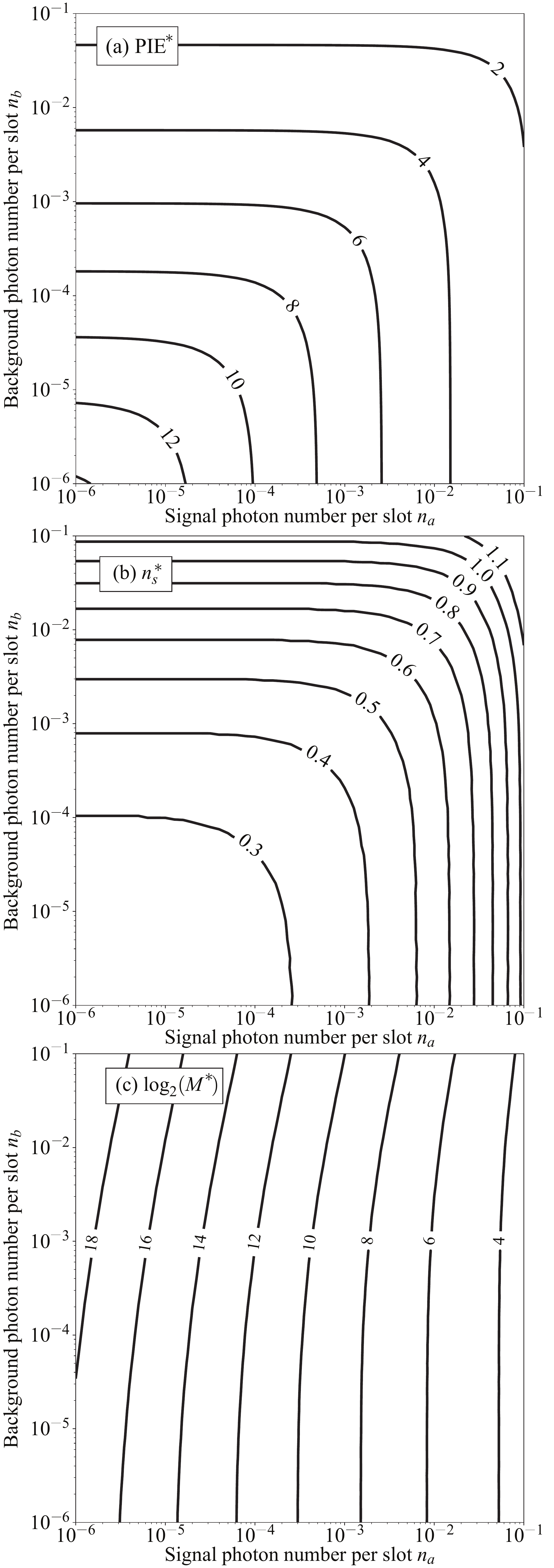}
\caption{(a) The maximum attainable photon information efficiency $\text{PIE}^\ast$ in bits/photon, (b) the corresponding optimal photon number per symbol $n_s^\ast$, and (c) base-2 logarithm of the optimal format order $\log_2 M^\ast$ as a function of the average detected numbers of signal photons $n_a$ and background photons $n_b$ per slot.}\label{Fig:Optimization}
\end{centering}
\end{figure}

\begin{figure}
\begin{centering}
\includegraphics[width=0.9\columnwidth]{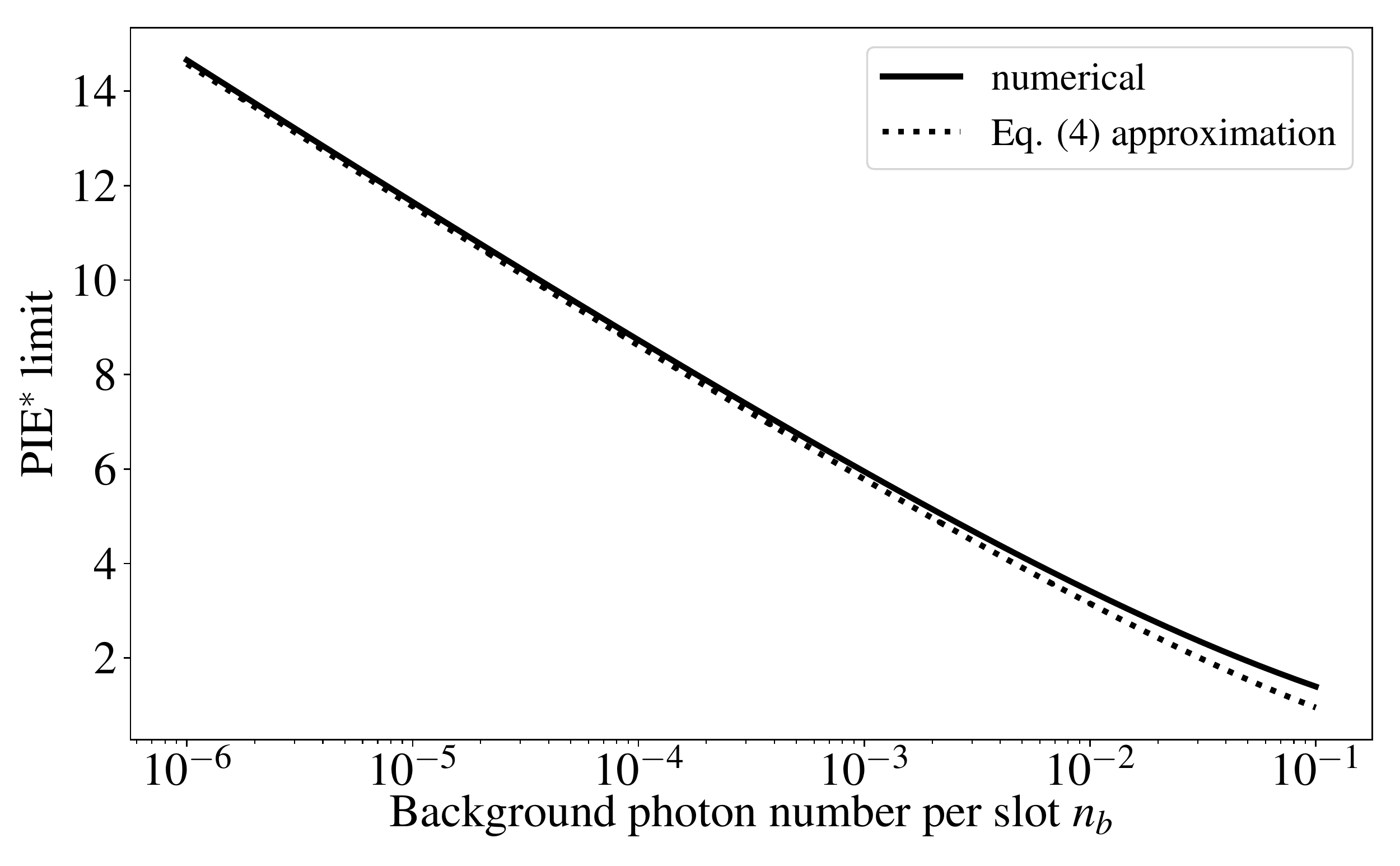}
\caption{Photon information efficiency in the limit of the vanishing signal strength $n_a \rightarrow 0$ as a function of the background photon number per slot $n_b$ calculated by numerical means (solid line) compared with the approximation given in Eq.~(\ref{Eq:PIEApprox}) (dotted line).}\label{Fig:PIEComparison}
\end{centering}
\end{figure}

For a given pair of the signal photon number $n_a$ and the background photon number $n_b$ per slot we optimized numerically the relative entropy bound over the format order $M$ to obtain the photon efficiency limit $\textrm{PIE}^\ast(n_a, n_b)$. Results of the optimization, carried out without any constraint on $M$, are shown in Fig.~\ref{Fig:Optimization}(a). It is seen that in the case of a substantial imbalance between the signal strength and the noise strength, the PIE limit depends predominantly on the higher of the two arguments. In the strongly negative SNR regime, when $n_a \ll n_b$, the PIE limit is effectively a function of the background noise strength only. Having borrowed mathematical tools from an earlier analysis of the noiseless scenario \cite{JarzynaKuszajOPEX2015}, we found that the limiting PIE value in bits/photon is well approximated by a mathematical formula
\begin{equation}
n_a \ll n_b: \quad \textrm{PIE}^\ast \approx \{ W(2/n_b) -2 + [W(2/n_b)]^{-1} \} \log_2\mathrm{e},
\label{Eq:PIEApprox}
\end{equation}
where $W(\,\cdot\,)$ is the Lambert $W$ function defined as a solution to the equation $x=W(x)\exp[W(x)]$.
For large arguments $x\gg 1$, the Lambert function has asymptotic expansion $W(x) \approx \log x - \log\log x$. This implies a weak, logarithmic dependence of the PIE limit on the actual background noise strength $n_b$.
The approximate formula given in Eq.~(\ref{Eq:PIEApprox}) is compared with the numerical result in Fig.~\ref{Fig:PIEComparison}.

\section{Discussion}

The optimal format order $M^\ast$ can be equivalently characterized by the symbol photon number $n_s^\ast = M^\ast n_a$ that maximizes PIE. This quantity is depicted in Fig.~\ref{Fig:Optimization}(b) as a function of the average detected signal and background photon numbers. As a rule of thumb, in the photon-starved regime the optimal symbol photon number $n_s^\ast$ should be kept just below one to ensure that the mode carrying the optical energy of the symbol is likely to generate a photocount. The optimal symbol duration
\begin{equation}
t_s^\ast = n_s^\ast \cdot
\left(\eta_{\textrm{rx}}\cdot \eta_{\textrm{ch}} \cdot \frac{P_{\textrm{tx}}}{hf_c}
\right)^{-1},
\end{equation}
needs to be long enough to carry $n_s^\ast$ photons given the detected signal photon flux and hence it scales with the link range as $t_s^\ast \propto r^2$.

When the modulation bandwidth $B$ is fixed, the average number $n_a$ of signal photons per slot scales as $r^{-2}$ with the distance. In order to maintain $n_s^\ast$ in the range of $0.1$--$1$ photons per symbol, the format order should increase correspondingly as $M^\ast=n_s^\ast \cdot n_a^{-1} \propto r^2$, resulting soon in dramatically high values shown Fig.~\ref{Fig:Optimization}(c). This leads to technical challenges, such as the growing peak-to-average power ratio of PPM signals. Furthermore, error correction becomes more problematic, as it needs to cope with multiple background counts within one symbol frame. An alternative approach is to lower the modulation bandwidth $B$ accordingly with the decreasing received signal power in order to keep $n_a$ at a fixed level. Provided that the optical spectrum of the signal is defined by the modulation bandwidth rather than e.g.\ by the linewidth of the transmitter master laser, the spectral acceptance window of the receiver can be adjusted in line with $B$ so that the background photon number $n_b$ stays at an approximately constant level.

In the variable-bandwidth scenario described above, the operating point $(n_a, n_b)$ remains fixed and so does the modulation format order $M^\ast$, while the slot duration $B^{-1}$ and the symbol duration $t_s^\ast = M^\ast \cdot B^{-1}$ scale as $r^2$. The choice of a mildly negative SNR operating point with $n_a \lesssim n_b$ offers comparatively good photon efficiency while reducing challenges stemming from the implementation of a very high order modulation format. An important design consideration for the error correcting code is the typical number of background counts expected within one symbol frame, characterized in the present case by $M^\ast n_b = n_s^\ast \cdot {\text {SNR}}^{-1}$.

Among the potential variety of scalable modulation formats, the recently suggested use of Hadamard BPSK words \cite{GuhaPRL2011,BanaszekICSOS2017} would shift the bulk of the complexity of implementing photon-efficient communication to the receiver subsystem, while utilizing a standard phase-modulation transmitter setup.
Furthermore, one could envisage dual-mode operation accommodating both coherent detection, with lower PIE but simpler realization, and photon-efficient direct detection of Hadamard words converted using a scalable interferometric setup into the PPM format.

The effects of atmospheric turbulence on phase shift keyed signals can be mitigated using multi-aperture terminals with digital postprocessing for coherent detection \cite{GeislerYarnallOPEX2016}, or multimode interferometric receivers \cite{SodnikICSOS2012,JinAgnePRA2018} tolerant to wavefront distortion in the case of photon counting. The analysis presented here applies as long as the symbol duration $t_s^\ast$ remains below the coherence time of the received signal. Also, the effect of detectors dark counts becomes more pronounced with the increasing symbol duration, but should remain minute given current advances in superconducting single photon detectors \cite{MarsiliVermaNPHOT2013,ShibataShimizuOL2015}.

\section*{Acknowledgment}

Insightful discussions with C. Heese, Ch. Marquardt, and M. D. Shaw are gratefully acknowledged.

\end{document}